\documentstyle[hip-artc]{article}
\volnumber{0}  \edyear{0000}  \frompage{000} \topage{000}                
\recrevdate{15 July 1999}

\authors{
{\twerm Yu.M.Sinyukov$^1$}\\[2.812mm]
{\normalsize
\hspace*{-8pt}$^1$ Bogolyubov Institute for Theoretical Physics, \\ 
252143 Kiev, Ukraine\\[0.2ex] 
}}
\abstract{
The single- and multi- particle inclusive spectra for strongly inhomogeneous
thermal boson systems are studied using the method of statistical operator.
The thermal Wick's theorem is generalized and the analytical solution of the
problem for an boost-invariant expanding boson gas is found. The results
demonstrate the effects of inhomogeneity for such a system: the spectra and
correlations for particles with wave-lengths larger than the system's
homogeneity lengths change essentially as compared with the results based on
the local Bose-Einstein thermal distributions. The effects noticeable grow
for overpopulated media, where the chemical potential associated with
violation of chemical equilibrium is large enough.}

\begin{document}

\title{Boson spectra and correlations for thermal locally equilibrium systems }
\author{}
\maketitle

\section{Introduction}

The theoretical study of thermalized hadron and quark-gluon systems is
important for understanding of the early Universe and new phenomena in the
current and future experiments at SPS, RHIC and LHC. The systems formed in
ultra-relativistic nucleus-nucleus collisions produce $10^3\div 10^5$
secondaries \cite{Satz}. The number of quarks and gluons exceeds this
estimate by one order of the value if the conditions for a phase transition
to the QCD-plasma are realized. It is reasonable to expect that this
quasi-macroscopic system could be thermalized during small proper time $\tau
_{0\mbox{ }}$ after initial collision \cite{McLerran}. The systems formed in
A+A collisions may be rather inhomogeneous ones even at the final decoupled
stage $\tau =\tau _f$ of their evolution because of a strong expansion and
very high initial density.

The current pion interferometry analysis at SPS CERN shows the effective
sizes of such systems are $R\simeq 3\div 7$ $fm$ \cite{Interf-Exper}. For a
homogeneous static source the value $R^2$ is its geometrical mean-square
size \cite{Kopylov}, for hydrodynamically expanding systems the longitudinal
interferometry size is approximately proportional to the hydrodynamic
length, $R_L\propto \lambda _{hydr}\equiv \left| v_{hydr,L}^{^{\prime
}}\right| ^{-1}\simeq \tau _f,$ in a central rapidity region \cite{Sin-qm88}%
. In general case the ''pion interferometry microscope'' measures the
lengths of homogeneity of hadron systems at the final stage. Naturally, for
earlier stages, $\tau \leq \tau _f$, the effective geometrical or
hydrodynamic lengths are smaller than the mentioned ones. At the initional
stage of thermalization the typical hydrodynamic length (longitudinal length
of homogeneity averaged over momenta ) is $\lambda _{hydr}\propto \tau _{0%
\mbox{ }}\simeq 1$ $fm$.

The statistical description of a quantum-field system with small homogeneity
regions should be done carefully. As well known the statistical
hydrodynamics can be based on the relativistic kinetic theory \cite{DeGroot}
as well as on the method of a nonequilibrium statistical operator \cite
{Zubarev}. In the both methods one uses the locally equilibrium distribution
as the zero approximation. Then the complete Wigner function is usually
represented by $f(x)=f_0(x,p)+\widetilde{f}(x,p)$ and nonequilibrium
statistical operator is $\rho (\sigma )={\bf \rho (}\sigma )+\widetilde{\rho 
}(\sigma ).$ Here $f_0(x,p)$ is a locally equilibrium distribution, the
quasi-equilibrium statistical operator ${\bf \rho (}\sigma )$ is defined on
some hypersurface $\sigma $ and corresponds to maximum entropy principle
under a given set of additional conditions on local averages such as the
energy density, the charge density, etc. The function $\widetilde{f}(x,p)$
and the operator $\widetilde{\rho }(\sigma )$ describe the nonequilibrium
flows associated with a heat, viscosity, etc. They give the contributions
that are roughly proportional to the ratio of the correlation length (mean
free path) to the hydrodynamic length. As to the main approximation, the
distribution function $f_0(x,p)$ is chosen usually in the form of globally
equilibrium distribution with the substitutions: $T\rightarrow T(x)\equiv
1/\beta (x),v\rightarrow v_{hydr}(x)$, etc., where the parameters depend now
on point $x$. This prescription is a physically self-consistent, if the
hydrodynamic length is much more than the Compton (or de Broglie for
massless fields) wavelengths $\lambda _p$ of the quanta, $\lambda _p\ll
\lambda _{hydr}$ \cite{DeGroot}. The condition cannot be satisfied for the
effective mass $m\leq 0.1$ $GeV$ at the early stage, $\tau \simeq \lambda
_{hydr}\approx 1fm$ , of the matter evolution in nucleus-nucleus collisions.
One of the aims of this paper is to find the locally equilibrium function $%
f_0(x,p)$ in the general case of an arbitrary ratio $\lambda _p$ to $\lambda
_{hydr}$. We shall use, for the purpose, the method of the quasi-equilibrium
statistical operator \cite{Zubarev}, \cite{Smolyansky}, \cite
{Panferov-ITP-Pr}.

The another problem, we discuss here, is how to calculate the two (many)-
particles inclusive spectra for very inhomogeneous locally equilibrium
systems. Strongly speaking, it is impossible to define the one-particle
Wigner function in this case: the necessary condition that the thermal
averages $<a^{\dagger }(p+\frac q2)a(p-\frac q2)>$ have to be diagonal
enough, i.e. q$_{eff}^2=\lambda _{hydr}^{-2}\ll m^2$\cite{DeGroot}, does not
satisfied$.$ In such a situation the direct use of the method of statistical
operator is appropriate for the spectra calculation.

The special interest for discussion is the appearance of the additional
terms in the double particle inclusive spectra connected with the non-zero
averages of the products of the creation and annihilation operators, $%
\left\langle a^{+}a^{+}\right\rangle $ and $\left\langle aa\right\rangle $.
Such terms have been obtained first for pions in the Gaussian current model 
\cite{Weiner}. The additional terms in the correlation function do not
appear in the nonrelativistic quantum-mechanical approach \cite{Sin-Tolst}.
We shall consider all these problems using the generalized Wick's theorem
for thermal locally equilibrium systems and give the explicit analytical
structure of the pair-correlation function for different sorts of bosons.

In the section 2 we deal with the description of inclusive spectra in the
standard Wigner's representation and discuss some basic points of this
approach.

In the section 3 we develop the statistical operator formalism for
calculation of bosonic operator averages, $\left\langle
a^{+}(p_1)a(p_2)\right\rangle $, $\left\langle a(p_1)a(p_2)\right\rangle $ ,
etc., in the locally equilibrium systems and discuss the physical conditions
for the hydrodynamic solutions we are interested in.

In the section 4 the thermal Wick's theorem is generalized for locally
equilibrium systems. This is the basis for the calculation of the double-
and multi- particle inclusive spectra.

The section 5 is devoted to the analytical calculation of the single
particle spectra for a physically important case of the boost-invariant
expansion of a hadron and/or quark-gluon matter. We derive there the
correction term to the Bose-Einstein heat spectrum and demonstrate the tie
of the term with the spectrum of the so-called Milne's particles.

In the section 6 we obtain the structure of the pion-, kaon-, and photon-
pair correlation functions. We calculate also the maximum value of the
interferometry peak for these particles and the analytical approximation for
the two-particle correlation function in typical experimental situations
when $m{}{}{}\tau _f$ $\gg 1$.

\section{The Statement of the Problem}

The description of the inclusive spectra and correlations for a
multiparticle production is based on a computation of the following type of
the averages

\begin{equation}  \label{spectra-def}
p^0\frac{dN}{d{\bf p}}=\left\langle a_p^{+}a_p\right\rangle ,\mbox{ }%
p_1^0p_2^0\frac{dN}{d{\bf p{\tt _1}}d{\bf p}{\tt _2}}=\left\langle
a_{p_1}^{+}a_{p_2}^{+}a_{p_1}a_{p_2}\right\rangle , \mbox{ etc,...}
\end{equation}

\noindent where $a_p^{+}$ and $a_p$ are the creation and annihilation
operators, corresponding to a quantum field $\varphi _{out}(x,t)$ when an
interaction is switched off. The brackets $<$...$>$ mean the average over
some density matrix describing the state of the system on a some
hypersurface $\sigma $. In the $S$-matrix theory the state is the $\left|
out\right\rangle $-state at $\sigma :t=\infty $. In the statistical
thermodynamic models of a multiparticle production the density matrix is
chosen to be the statistical operator ${\bf \rho }$ and surface $\sigma $ is
usually a freeze out hypersurface. The averages (\ref{spectra-def}) taken on
this hypersurface are coincided approximately with ones taken on an
arbitrary hypersurface that situated within of the light cone of the future
as to the freeze-out hypersurface. It corresponds to the preservation of the
momentum distributions of free streaming particles in Eqs.(\ref{spectra-def}%
) if one neglects the final state interaction and Coulomb corrections. The
hypersurface $\sigma $ can correspond also to an earlier stage of the
evolution if one studies the dilepton or photon productions from a hadron
and/or quark-gluon plasma. It the latter case the operators $a_p^{+}$ and $%
a_p$ have to correspond to weekly interacting quasi-particles with the
standard relativistic form of the dispersion relations in the medium and the
masses depend now on the temperature and density. In this paper we shall not
consider such a situation in details.

The inclusive double particle spectrum in (\ref{spectra-def}) is usually
calculated under supposition that the four-operator averages can be
decomposed into the products of the irreducible two-operator ones

\begin{equation}  \label{operator-decomposition}
\left\langle a_{p_1}^{+}a_{p_2}^{+}a_{p_1}a_{p_2}\right\rangle =\left\langle
a_{p_1}^{+}a_{p_1}\right\rangle \left\langle a_{p_2}^{+}a_{p_2}\right\rangle
+\left\langle a_{p_1}^{+}a_{p_2}\right\rangle \left\langle
a_{p_2}^{+}a_{p_1}\right\rangle
\end{equation}

\noindent In this case the problem of the inclusive multi-particle spectra
is reduced to the calculation of the $\left\langle
a_{p_1}^{+}a_{p_2}\right\rangle $ averages.

If one considers the free identical particles 1 and 2 the two-operator
average can be expressed by means of the following distribution function

\begin{equation}  \label{wigner-def}
f(x,p)=(2\pi )^{-3}\int d^4u\delta (p\cdot u)e^{-iu\cdot x}\left\langle
a^{+}(p-\frac u2)\;a(p+\frac u2)\right\rangle
\end{equation}

\noindent where $p=(p_1+p_2)/2$ and the average is done in a space-time
region where the interaction is negligible. Indeed, if we consider some
hypersurface $\Sigma $ that situated in this region and can be closed by a
plane surface $t=const$, it is possible to use the equation

\begin{equation}  \label{Gauss}
\int d\Sigma _\mu p^\mu e^{ik\cdot x}=(2\pi )^3p^0e^{ik^0t}\delta ^3({\bf k)}%
\mbox{ \qquad if \quad }p\cdot k=0
\end{equation}

\noindent that follows from the Gauss theorem. Then

\begin{equation}  \label{average-wigner}
\begin{array}{l}
\left\langle a^{+}(p_1)a(p_2)\right\rangle =\int
d^4up^0e^{i(q^0-u^0)t}\delta (p\cdot u)\delta ^3( {\bf q-u)}\left\langle
a^{+}(p-\frac u2)\;a(p+\frac u2)\right\rangle \\ 
\\ 
=\int d\sigma _\mu p^\mu e^{iq\cdot x}f(x,p)
\end{array}
\end{equation}

\noindent Here $\sigma $ is the part of the hypersurface $\sum $ where $%
f(x,p)\neq 0$, $q=p_2-p_1.$

The expression (\ref{average-wigner}) is general and describes the
operator's averages for the radiating matter when the hypersurface $\sigma $
is an arbitrary hypersurface situated within of the light cone of the future
as to the decoupling 4-volume. In the general case the function $f(x,p)$ is
rather complicated, even not positively defined. The distribution function $%
f(x,p)$ is coincided with the single- particle Wigner function $f_W(x,p)$
for free fields if $<a^{\dagger }(p+\frac q2)a(p-\frac q2)>$ is diagonal
enough: $q_{eff}^2\ll m^2.$ Then there is the direct tie between the
distribution function (\ref{wigner-def}) and complete Wigner function $%
N(x,p)=\frac 124\pi (2\pi )^{-5}\int d^4v\exp (-ipv)<:\varphi (x+\frac
12v)\varphi (x-\frac 12v):>$ \cite{DeGroot}.

The real calculations of the final spectra and correlations simplify greatly
if a system is thermal and decoupling volume is narrow enough in time-like
direction and so it can be considered as the freeze-out hypersurface. In
this case one can use the thermal matrix density ${\bf \rho }$ at this
freeze-out hypersurface $\sigma $ and calculate the phase-space distribution
function $f(x,p)$ directly. When the surface $\sigma $ changes from event to
event it is necessary to do the additional average over $\sigma $ in the all
final expressions for spectra. But this procedure cannot be used even
formally for description of two ( many)-particle spectra when the radiation
volume is an essentially 4-dimensional one. For this aim instead of single
particle Wigner function (let us suppose here that $f(x,p)=f_W(x,p)$) one
have to use the quasi-classical density of particle emission $%
g(x,p)=p^0d^7N/d^4xd^3p$. The latter can be expressed by means of the
derivation of complete Wigner function, $p^\mu \partial _\mu N(x,p)$, that
take into account the interaction in the system leading to continuous
radiation during some finite time. The decay of the resonances is one of an
example of such a 4-volume emission. We will not consider this case in the
paper.

As we show hereinafter, it is convenient to choose the one-component scalar
field as the basic model of our consideration. The Lagrange function has the
form {\sl \ {\cal L}}={\cal L}{\sl $_{KG}+${\cal L}$_{INT}$} corresponding
to the free Klein-Gordon field and an interaction term. If the latter is
characterized by the coupling constant $\alpha $ (with the dimension equal
to product of energy and volume), we can neglect the interaction energy and
momentum if the following conditions for temperature $T$ and particle
density $n(x)$ are satisfied \cite{DeGroot}

\begin{equation}  \label{neglect-int}
\frac{n\alpha }T<<1
\end{equation}

Neglecting the interaction terms, the single-particle Wigner function near
the mass-shell is associated with the local current \cite{DeGroot}. To be
simple we will consider here the real field and use the particle flow as the
current 
\begin{equation}
j_\mu (x)=\varphi ^{(+)}(x)\stackrel{\leftrightarrow }{\frac \partial
{\partial x^\mu }}\varphi ^{(-)}(x)  \label{j(x)}
\end{equation}

\noindent where the decomposition of the field into ''positive'' and
''negative'' parts looks like 
\begin{equation}
\varphi (x)=\varphi ^{(+)}(x)+\varphi ^{(-)}(x)\equiv [2(2\pi )^3]^{-\frac
12}\int \frac{d^3p}{p^0}\left[ a_p^{+}e^{ip\cdot x}+a_pe^{-ip\cdot x}\right]
\label{field-phi}
\end{equation}

The expressions for the single particle spectrum and the double particle
correlation function follow immediately from Eqs. (\ref{spectra-def}),(\ref
{operator-decomposition}),(\ref{average-wigner}):

\begin{equation}  \label{single1}
p^0\frac{dN}{d{\bf p}}=\int d\sigma _\mu p^\mu f(x,p)
\end{equation}

\begin{equation}  \label{double1}
C(p_1,p_2)=1+\left( p_1^0p_2^0\frac{dN}{d{\bf p_1}}\frac{dN}{d{\bf p_2}}%
\right) ^{-1}\left| \int d\sigma _\mu p^\mu e^{iq\cdot x}f(x,p)\right| ^2
\end{equation}

If one considers the infinite homogeneous thermodynamic system (with the
chemical potential $\mu $, the energy-momentum operator $\widehat{P}$ and
the operator of particle number $\widehat{N}$) that moves as a single whole
with 4-velocity $u^\mu $, the result of the averaging over the equilibrium
statistical operator

\begin{equation}
{\bf \rho }_{eq}=\frac 1Z\exp \left[ \left( -\widehat{P}^\nu u_\nu +\mu 
\widehat{N}\right) /T\right]  \label{rho-equilibr}
\end{equation}

\noindent is the following \cite{DeGroot} 
\begin{equation}
\left\langle a^{+}(p_1)a(p_2)\right\rangle =(2\pi )^3p^0f_{B.E.}(p)\delta ^3(%
{\bf p}_1-{\bf p}_2)\mbox{, }\left\langle a(p_1)a(p_2)\right\rangle =0
\label{inhomog-average}
\end{equation}

\noindent where $f_{B.E.}(p)$ is the Bose-Einstein distribution for the
globally-equilibrium systems. Let us put for simplicity $\mu =0$. Then 
\begin{equation}
f(p,x)=\frac{(2\pi )^{-3}}{\exp (\beta p\cdot u)-1}\equiv f_{B.E.}(p;\beta
,u);\,\beta =\frac 1T=const,u^\nu =const.  \label{BE-distr}
\end{equation}

Here the Wigner function does not depend on $x$. The thermal Wick's theorem
can be proved for a such type of systems, that leads to the result (\ref
{operator-decomposition}). The main approximation for the Wigner function of
an expanding hadron and quark-gluon gas is usually based on the
distributions like (\ref{BE-distr}) with the substitutions $\beta
=const\rightarrow \beta (x)$, $u=const\rightarrow u(x).$ Such substitutions
are physically reasonable if the wavelength of the corresponding quanta, $%
\lambda _p=1/m_{eff}$, is much less than the hydrodynamic length

\begin{equation}  \label{unequality}
\lambda _p<<\lambda _{hydr}\propto \min \{\left| \partial ^\mu u_\mu
^{*}\right| ^{-1},\left| gradT\right| ^{-1}\}
\end{equation}

Asterisk marks the values in the (local) rest system. As it was mentioned in
Sec.1, the typical hydrodynamic length is approximately equal to the proper
time of the hydrodynamic expansion, $\tau _f$ , and the inequality (\ref
{unequality}) is satisfied if $m\tau \gg $ $1$. For pions and kaons $m\tau
_f\geq $1, for chiral quarks and gluons $m\tau \ll 1$. For thermalized
photons there is always the momentum region where inequality (\ref
{unequality}) is strongly violated.

At the end of this section we would like to emphasize that the results (\ref
{inhomog-average}), (\ref{BE-distr}) for averages $\left\langle
a_{p_1}^{+}a_{p_2}\right\rangle $ as well as the zero values for the
averages $\left\langle a_{p_1}^{+}a_{p_2}^{+}\right\rangle $ have been
derived for infinite homogeneous equilibrium systems only and cannot be
automatically applied to locally equilibrium inhomogeneous system by using
the simple substitution $\beta =const\rightarrow \beta (x)$, $%
u=const\rightarrow u(x)$ in the Bose-Einstein distribution (\ref{BE-distr}).
This concerns also of the two-particle spectra (\ref{double1}) and expansion
(\ref{operator-decomposition}) which is based on the Wick's theorem for
globally equilibrium systems.

It is interesting to mention that despite of the different structure of the
correlation function (\ref{double1}) in different approaches (e.g.,\cite
{Pratt-86,Kolehmanien,Hama,Makhl-Sin,Weiner}) they will give approximately
same results being applied to systems that are quasi-homogeneous ones, $%
m\tau _f\gg 1,$ and are described by the same Wigner functions. But for
strongly inhomogeneous systems it is impossible to introduce by a standard
way the single particle Wigner function as well as to preserve the structure
of the correlation function (\ref{double1}) based on Eq.(\ref
{operator-decomposition}). Therefore, all these approaches are off the
region of their applicability. One have to calculate the averages such as $%
\left\langle a_{p_1}^{+}a_{p_2}\right\rangle $, $\left\langle
a_{p_1}^{+}a_{p_2}^{+}\right\rangle $ directly. The formal distribution
functions defined by (\ref{wigner-def}) will be differ from local
Bose-Einstein distribution even for ideal Bose gas. The structure (\ref
{double1}) of the correlation function will be destroyed altogether with the
Bose-Einstein distribution (\ref{BE-distr}). In the following sections we
propose the method for study of spectra and correlations in inhomogeneous
thermal systems.

\section{The Method of Locally Equilibrium Statistical Operator}

The hydrodynamic description of quantum-field system, as known, can be based
on the method of non-equilibrium statistical operator \cite{Zubarev}, \cite
{Smolyansky}, \cite{Panferov-ITP-Pr}, \cite{VonWeert}. The initial step in
this method is to construct the so-called quasi-equilibrium statistical
operator that describes hydrodynamics of the system neglecting the viscosity
effects, heat conductivity, etc. In other words, the operator describes the
locally equilibrium systems and since it will be used for this aim only we
will call it as the locally equilibrium operator ${\bf \rho }$. To build the
operator one usually applies the maximum entropy principle \cite{Zubarev},
\cite{VonWeert}. The method carries into effect in a full analogy with the
Gibbs method for homogenous equilibrium systems. In the latter case the set
of the averages $<\widehat{E}>,$ $<\widehat{P}>,$ $<\widehat{Q}>,$ etc., is
considered as the fixed additional conditions when the entropy $S$ is
maximized. For the locally equilibrium systems considered on some
hypersurface $\sigma $ with a time-like normal vector $n^\nu $ the
collection of the additional conditions is based on densities of energy $%
\varepsilon (x)$, momentum ${\bf p}(x)$, charge $q(x)$, etc. In the
relativistic covariant form they look like

\begin{equation}
\left\langle n_\nu (x)\widehat{T}^{\mu \nu }(x)\right\rangle ,\quad
\left\langle n_\nu (x)\widehat{J}^\nu (x)\right\rangle
\label{additional-cond}
\end{equation}

\noindent where $\widehat{T}^{\mu \nu }(x)$ is the operator of the
energy-momentum tensor, $\widehat{J}^\nu (x)$ is current density operator.
According to the definition

\begin{equation}  \label{Rho-S}
{\bf \rho =}e^{-S(\sigma )}
\end{equation}

The entropy is maximized under the additional conditions like (\ref
{additional-cond}) by the Lagrange multipliers method

\begin{equation}
S=\max \;Sp\left[ -{\bf \rho }\ln {\bf \rho -\rho }\int d\sigma \;n_\gamma
(x)(\beta _\nu \widehat{T}^{\nu \gamma }(x)\;\;-\mu \beta \widehat{J}^\gamma
(x))-\lambda {\bf \rho }\right]  \label{Lagrange-method}
\end{equation}

\noindent where Lagrange multiplier $\beta _\nu (x)=u^\nu (x)/T(x)$ \cite
{VonWeert}. The formal solution of Eq.(\ref{Lagrange-method}) gives us the
result for entropy \cite{Panferov-ITP-Pr}, \cite{VonWeert}

\begin{equation}
S=S(\sigma )=\Phi (\sigma )+\int d\sigma \;n_\gamma (x)(\beta _\nu \widehat{T%
}^{\nu \gamma }(x)-\mu \beta \widehat{J}^\gamma (x))  \label{S}
\end{equation}

\noindent where $\Phi (\sigma )=\ln \;Sp\exp \{\int d\sigma \;n_\gamma
(x)(\beta _\nu \widehat{T}^{\nu \gamma }(x)-\mu \beta \widehat{J}^\gamma
(x))\}$ is Masier-Planck functional. Let us put $\mu =0$ for simplicity.
Sticking to the analogy with the method of statistical operator for globally
equilibrium systems where all the operators under the sign $Sp$ are mutually
commuted, we demand that

\begin{equation}  \label{commutation-Sp}
\;\left[ \;n_\nu (x)\beta _\mu (x)\widehat{T}^{\mu \nu }(x)\;,\;\;n_\gamma
(y)\beta _\delta (y)\widehat{T}^{\delta \gamma }(y)\right] _\sigma =0
\end{equation}

In this case the set of the additional conditions (\ref{additional-cond})
has the standard interpretation and we hope to avoid some mathematical
difficulties that may possibly appear in a more general case.

As we mentioned in Sec.2, we shall start from the free one-component scalar
field (\ref{field-phi}) with the standard commutation relations for the
operators

\begin{equation}  \label{commut-relation}
\left[ a(p),a^{+}(p^{\prime })\right] =p^0\delta ^3({\bf p-p}^{\prime }{\bf )%
}
\end{equation}

\noindent The energy-momentum tensor has the form

\begin{equation}  \label{tensor}
\widehat{T}^{\mu \nu }(x)=\frac{\partial \varphi }{\partial x_\mu }\frac{%
\partial \varphi }{\partial x_\nu }-g^{\mu \nu }{\cal L_{{\rm KG}}}
\end{equation}

The commutation equation (\ref{commutation-Sp}) is solved using the Eqs.(\ref
{field-phi}), (\ref{commut-relation}), (\ref{tensor}) and gives the
following conditions for hydrodynamic values taken on a hypersurface $\sigma
:$%
\begin{equation}  \label{u-n-x-conditions}
\begin{array}{l}
n^\mu (x)=u^\mu (x) \\ 
\\ 
x\cdot u(x)=y\cdot u(y)\qquad \forall x,y\in \sigma
\end{array}
\end{equation}

It immediately follows from the Eqs.(\ref{u-n-x-conditions}) that the
commutation relation (\ref{commutation-Sp}) is satisfied when one of the
following conditions is realized:

1. The hypersurface $\sigma $ is plane and $u^\mu (x)=const$ ($v(x)=0$ in
the reference system where $\sigma $ is $t=const$ ) . Actually, this means
that the system occupies the space-time region with some distribution in
temperature (falling down from central highly excited part to the vacuum at
the periphery) and has no internal motion.

2. The hypersurface $\sigma $ is defined by the condition $t^2-{\bf x}%
_L^2=\tau ^2=const$, and the one-dimensional expansion along $L$-axis occurs
with 4-velocity $u_0=t/\tau $, $u_L=x_L/\tau ,$ ${\bf u}_T=0.$ If $\beta $
is a constant on $\sigma $, it is reduced to the well-known boost-invariant
expansion \cite{Gor-Sin-Zhd}, \cite{Bjorken}, that is the basic model for
application of the hydrodynamic theory to multiple processes in high energy
collisions.

3. The hypersurface $\sigma $ is defined by the condition $t^2-{\bf x}%
_T^2=\tau ^2=const$ and there holds the two-dimensional expansion $%
u_0=t/\tau $, ${\bf u}_T=x_T/\tau ,$ $u_L=0$.

4. The hypersurface $\sigma $ is defined by the condition $t^2-{\bf x}%
^2=\tau ^2=const$, 3-dimensional hydrodynamical expansion has the form $%
u^\mu =x^\mu /\tau $.

The method proposed to be used to find the averages of the operator products
is based on the Gaudin's idea \cite{Gaudin} for globally equilibrium systems
and is the following. We represent the locally-equilibrium statistical
operator ${\bf \rho }$ defined by (\ref{Rho-S}), (\ref{S}) in the form 
\begin{equation}
{\bf \rho =}\frac{1}{Z}\exp \left[ {\bf -}\int d\sigma _{\nu }\beta _{\mu }%
\widehat{T}^{\mu \nu }(x)\right] \qquad   \label{rho-alpha}
\end{equation}

\noindent where the integral is taken over corresponding hypersurface $%
\sigma $ as it was discussed before. Let us introduce the operators that
dependent on some  parameter $\alpha $ in the following way

\begin{equation}  \label{a-p-alpha}
\begin{array}{l}
a^{+}(p,\alpha )=e^{{\bf -}\alpha \int d\sigma _\nu \beta _\mu \widehat{T}%
^{\mu \nu }(x)}a^{+}(p)\;e^{\alpha \int d\sigma _\nu \beta _\mu \widehat{T}%
^{\mu \nu }(x)} \\ 
\\ 
a(p,\alpha )=e^{{\bf -}\alpha \int d\sigma _\nu \beta _\mu \widehat{T}^{\mu
\nu }(x)}a(p)\;e^{\alpha \int d\sigma _\nu \beta _\mu \widehat{T}^{\mu \nu
}(x)}
\end{array}
\end{equation}

\noindent and use the matrix notation

\begin{equation}  \label{A-p-alpha}
{\bf A(}p,\alpha )\equiv \left( 
\begin{array}{c}
a^{+}(p,\alpha ) \\ 
a(p,\alpha )
\end{array}
\right) ,\qquad {\bf A(}p)\equiv \left( 
\begin{array}{c}
a^{+}(p) \\ 
a(p)
\end{array}
\right)
\end{equation}

\noindent It is easy to get the following equations

\begin{equation}  \label{A-p-alpha-0}
{\bf A(}p,\alpha =0)={\bf A(}p)
\end{equation}
\begin{equation}  \label{A-p-alpha-1}
\begin{array}{c}
< {\bf A(}p)a(p^{\prime })>=<a(p^{\prime }){\bf A(}p,\alpha =1)>\; \\ 
\\ 
<{\bf A(}p)a^{+}(p^{\prime })>=<a^{+}(p^{\prime }){\bf A(}p,\alpha =1)>
\end{array}
\end{equation}

The latter equations follow from the trace invariance under the cyclic
permutation of operators. To express the operators ${\bf A(}p,\alpha )$
through $a^{+}(p)$ and $a(p)$ we shall use the equations that stem directly
from Eq.(\ref{a-p-alpha})

\begin{equation}  \label{diff-A-alpha}
\frac{\partial {\bf A(}p,\alpha )}{\partial \alpha }=e^{{\bf -}\alpha \int
d\sigma _\nu \beta _\mu \widehat{T}^{\mu \nu }(x)}\left[ {\bf A(}p)\;,\;\int
d\sigma _\nu \beta _\mu \widehat{T}^{\mu \nu }(x)\right] e^{{\bf -}\alpha
\int d\sigma _\nu \beta _\mu \widehat{T}^{\mu \nu }(x)}
\end{equation}

Using the commutator (\ref{commut-relation}) and conditions (\ref
{u-n-x-conditions}) we find the concrete form of Eq.(\ref{diff-A-alpha}) for
scalar field (\ref{field-phi})

\begin{equation}  \label{diff-A-alpha-K}
\frac{\partial {\bf A(}p,\alpha )}{\partial \alpha }=\int d^3k{\bf K}(p,k)%
{\bf A}(k,\alpha )
\end{equation}

\noindent where the matrix kernel of integro-differential equation (\ref
{diff-A-alpha-K}) has the form (asterisk means the complex conjugation) 
\begin{equation}  \label{matrix-K}
{\bf K}(p,k)=\left( 
\begin{array}{cc}
G(p,k) & \overline{G}(p,k) \\ 
-\overline{G}^{*}(p,k) & -G^{*}(p,k)
\end{array}
\right)
\end{equation}

\noindent where

\begin{equation}  \label{G}
\begin{array}{l}
G(p,k)=-\frac 1{(2\pi )^3}\int d\sigma \;e^{i(k-p)\cdot x} \frac{\beta (x)}{%
k^0}\left[ k\cdot u\;p\cdot u-(k\cdot p-m^2)/2\right] \\ 
\\ 
\overline{G}(p,k)=\frac 1{(2\pi )^3}\int d\sigma \;e^{-i(k+p)\cdot x}\frac{%
\beta (x)}{k^0}\left[ k\cdot u\;p\cdot u-(k\cdot p+m^2)/2\right]
\end{array}
\end{equation}

\noindent Here $\beta (x)$ is the inverse of the local temperature $T(x)$, $%
u(x)=n(x)$ is the hydrodynamic 4-velocity and we use here the integral
measure in the form $d\sigma _\mu =d\sigma \,n_\mu (x).$

The solution of the system of integro-differential equations (\ref
{diff-A-alpha-K}), which is considered according to Eq.(\ref{A-p-alpha-0})
as the Cauchy problem at ${\bf A(}p,\alpha =0)={\bf A(}p)$, is (see Ref.\cite
{Bykov}):

\begin{equation}
{\bf A(}p,\alpha )={\bf A(}p)+\sum_{n=1}^\infty \frac{\alpha ^n}{n!}\int d^3k%
{\bf K}_n(p,k){\bf A(}k)  \label{sol-A-p-alpha}
\end{equation}

\noindent where ${\bf K}_n(p,k)$ is the $n$-th iteration of matrix kernel $%
{\bf K}$:

\begin{equation}  \label{K-n}
\begin{array}{l}
{\bf K}_1(p,k)={\bf K}(p,k) \\ 
\\ 
{\bf K}_n(p,k)=\int ds_1ds_2...ds_{n-1}{\bf K}(p,s_1){\bf K}(s_1,s_2)...{\bf %
K}(s_{n-1},k)
\end{array}
\end{equation}

\noindent Using Eq.(\ref{matrix-K}) we have the properties for the
iterations of the kernel 
\begin{equation}  \label{K-n-properties}
{\rm K_n^{22}}=(-1)^n{\rm K}_n^{*11},\quad {\rm K_n^{21}}=(-1)^n{\rm K}%
_n^{*12}
\end{equation}

The system of integral equations for the operator averages follows from the
solution (\ref{sol-A-p-alpha}) of the integro-differential equation (\ref
{diff-A-alpha-K}) and the relation for averages (\ref{A-p-alpha-1}): 
\begin{equation}
\begin{array}{l}
\sum_{n=1}^\infty \frac 1{n!}\int d^3k\left( {\rm K}_n^{11}(p,k)\left\langle
a(p^{\prime })a^{+}(k)\right\rangle +{\rm K}_n^{12}(p,k)\left\langle
a(p^{\prime })a(k)\right\rangle \right) =-p^0\delta ^3(p-p^{\prime }) \\ 
\\ 
\sum_{n=1}^\infty \frac 1{n!}\int d^3k\left( {\rm K}_n^{21}(p,k)\left\langle
a(p^{\prime })a^{+}(k)\right\rangle +{\rm K}_n^{22}(p,k)\left\langle
a(p^{\prime })a(k)\right\rangle \right) =0
\end{array}
\label{integral-eqs1}
\end{equation}

\noindent and

\begin{equation}
\begin{array}{l}
\sum_{n=1}^\infty \frac 1{n!}\int d^3k\left( {\rm K}_n^{11}(p,k)\left\langle
a^{+}(p^{\prime })a^{+}(k)\right\rangle +{\rm K}_n^{12}(p,k)\left\langle
a^{+}(p^{\prime })a(k)\right\rangle \right) =0 \\ 
\\ 
\sum_{n=1}^\infty \frac 1{n!}\int d^3k\left( {\rm K}_n^{21}(p,k)\left\langle
a^{+}(p^{\prime })a^{+}(k)\right\rangle +{\rm K}_n^{22}(p,k)\left\langle
a^{+}(p^{\prime })a(k)\right\rangle \right) =p^0\delta ^3(p-p^{\prime })
\end{array}
\label{integral-eqs2}
\end{equation}

The integral equations (\ref{integral-eqs1}) and (\ref{integral-eqs2})
contain the complete information about the one- and many-particle inclusive
spectra for the locally equilibrium thermalized Klein-Gordon field. If the
system is an infinite homogeneous ($\beta $=const ) one and is considered on
a flat hypersurface $\sigma $: $t_{\ast }=const$ in the rest frame where $%
u_{\ast }^{\mu }$ $=const$, it immediately follows from (\ref{G}), (\ref{K-n}%
) that ${\rm K}^{11}=-\beta p_{\ast }^{0}\delta ^{3}({\bf p^{\ast }-k^{\ast
})}$, ($p_{0}^{\ast }=p\cdot u),$ and according to Eqs. (\ref{integral-eqs1}%
) and (\ref{integral-eqs2}) we have\noindent 
\begin{equation}
\left\langle a^{+}(p)a^{+}(p^{^{\prime }})\right\rangle _{eq}=\left\langle
a(p)a(p^{\prime })\right\rangle _{eq}=0,\,\left\langle a^{+}(p)a(p^{^{\prime
}})\right\rangle _{eq}=\frac{p^{0}\delta ^{3}({\bf p-p^{^{\prime }})}}{\exp
(\beta p\cdot u)-1}  \label{homog-average}
\end{equation}

This corresponds to the standard result (\ref{inhomog-average}) for globally
equilibrium systems and leads to the Bose-Einstein distribution for a
homogeneous ideal gas. In all other cases the result will be different from (%
\ref{homog-average}); the averages $\left\langle a^{+}a^{+}\right\rangle $
and $\left\langle aa\right\rangle $ do not vanish because $c$-factors
attached to the corresponding operator pairs in the tensor $T^{\mu \nu }$ do
not become zero after the integration over $\sigma $. Hereinafter we shall
consider the solution of Eqs.(\ref{integral-eqs1}) and (\ref{integral-eqs2})
for the concrete locally equilibrium systems.

\section{The Thermal Wick's Theorem for Locally Equilibrium Systems}

The double and multi-particles inclusive spectra are defined by Eqs.(\ref
{spectra-def}) and are expressed through the four and many operator
averages. If a system is in globally equilibrium state, there can be used
the thermal Wick's theorem to express the many-point operator averages as
the products of two point ones. For 4-point average the corresponding
expansion is given by the Eq.(\ref{operator-decomposition}). Our task now is
to generalize thermal Wick's theorem for locally equilibrium systems.

Let us consider the 4-point operator averages. First we introduce the
notation for the operator expression

\begin{equation}
\sum_{n=1}^\infty \frac 1{n!}\int d^3k{\rm K}_n(p,k)=\widehat{K}(p,k)
\label{notation}
\end{equation}

\noindent in order to simplify computation. Then the integral equations (\ref
{integral-eqs1}) and (\ref{integral-eqs2}) take the compact form 
\begin{equation}  \label{integral-eqs-compact}
\begin{array}{c}
\widehat{K}(p,k)\left\langle a(p^{\prime }){\bf A(}k)\right\rangle =\left( 
\begin{array}{c}
-p^0\delta ^3( {\bf p-p^{\prime })} \\ 
0
\end{array}
\right) \\ 
\\ 
\quad \widehat{K}(p,k)\left\langle a^{+}(p^{\prime }){\bf A(}k)\right\rangle
=\left( 
\begin{array}{c}
0 \\ 
p^0\delta ^3({\bf p-p^{\prime })}
\end{array}
\right)
\end{array}
\end{equation}

Using the trace invariance under the cyclic permutation and solution (\ref
{sol-A-p-alpha}) one can get convinced that 
\begin{equation}
\begin{array}{l}
\left\langle {\bf A(}p_1)a^{+}(p_2)a(p_1^{\prime })a(p_2^{\prime
})\right\rangle =\left\langle a^{+}(p_2)a(p_1^{\prime })a(p_2^{\prime }){\bf %
A(}p_1,\alpha =1)\right\rangle = \\ 
\\ 
\left\langle a^{+}(p_2)a(p_1^{\prime })a(p_2^{\prime }){\bf A(}%
p_1)\right\rangle +\widehat{K}(p_1,k)\left\langle a^{+}(p_2)a(p_1^{\prime
})a(p_2^{\prime }){\bf A(}k)\right\rangle
\end{array}
\label{Wick-Aaaa}
\end{equation}
After the commutation of the vector ${\bf A}${\bf \ }and the representation
of the arising $\delta $-function by means of the Eq.(\ref
{integral-eqs-compact}) we have 
\begin{equation}
\begin{array}{l}
\left\langle {\bf A(}p_1)a^{+}(p_2)a(p_1^{\prime })a(p_2^{\prime
})\right\rangle -\left\langle a^{+}(p_2)a(p_1^{\prime })a(p_2^{\prime }){\bf %
A(}p_1)\right\rangle = \\ 
\\ 
\widehat{K}(p_1,k){\tt [}\left\langle a(p_1^{\prime }){\bf A(}%
k)\right\rangle \left\langle a^{+}(p_2)a(p_2^{\prime })\right\rangle
+\left\langle a^{+}(p_2){\bf A(}k)\right\rangle \left\langle a(p_1^{\prime
})a(p_2^{\prime })\right\rangle + \\ 
\\ 
\left\langle a(p_2^{\prime }){\bf A(}k)\right\rangle \left\langle
a^{+}(p_2)a(p_1^{\prime })\right\rangle ]=\widehat{K}(p_1,k)\left\langle
a^{+}(p_2)a(p_1^{\prime })a(p_2^{\prime }){\bf A(}k)\right\rangle
\end{array}
\label{Wick-A-commutation}
\end{equation}

Just in a similar way one can derive the analogous equation using the
substitution $a^{+}(p_2)\rightarrow a(p_2)$ and the Hermitian conjugated to
it. It is also worthy to mention that the equality between the last two
parts of Eq.(\ref{Wick-A-commutation}) is preserved when the vector ${\bf A}$
commutes to the left side in all the brackets. The Eq.(\ref
{Wick-A-commutation}) is easy generalized for the case of any even number of
the operators by the induction method. All that means that the following
equation is valid for even number of operators 
\begin{equation}
\widehat{K}(p_1,k){\bf \Delta }\left( {\bf A(k),}A(p_2),...,A(p_j),...%
\right) ={\bf 0}  \label{Wick-K-A-0}
\end{equation}

\noindent where 
\begin{equation}
{\bf \Delta }=\left\langle {\bf A(k)}A(p_2)...A(p_j)...\right\rangle -\sum_{%
{\cal P}}\left\{ \left\langle {\bf A(k)}A(p_{j_k})\right\rangle
\prod\limits_{j^{\prime }>\,j;j,\;j,\prime \neq j_k}\left\langle
A(p_jA(p_{j^{/}})\right\rangle \right\}  \label{Wick-delta}
\end{equation}

\noindent and $A(p)$ is $a^{+}(p)$ or $a(p)$ and ${\cal P}${\cal \ }is the
permutation sign. If the integral operator $\widehat{K}$ is a nondegenerate
one, the Eq. (\ref{Wick-delta}) has the unique solution ${\bf \Delta =0}$
that means the average of any even number of operators expands in the sum of
the products of all operator pairs taken in the same order as they were in
the initial expression:

\begin{equation}
\left\langle A(p_1)A(p_2)...A(p_n)\right\rangle =\sum_{{\cal P}%
}\prod\limits_{j^{\prime }>\,j}\left\langle A(p_jA(p_{j^{\prime
}})\right\rangle ,\qquad (n=2k)  \label{Wick-theorem}
\end{equation}
\qquad

It is obvious that the averages of odd numbers of the operators are equal to
zero because of the bilinearity of the energy-momentum tensor $T^{\mu \nu
}(x)$ in operators $a^{+},a$. So, the theorem is proved. The main
peculiarity as compared with the standard results is the presence of
additional terms like $\left\langle a^{+}(p_1)a^{+}(p_2)\right\rangle $ and $%
\left\langle a(p_1)a(p_2)\right\rangle $ in the expansion. As it will be
shown in the next section these non-zero terms arise because of a space-time
finiteness of the homogeneity regions in locally equilibrium systems.

\section{Boson Spectra in the Boost-Invariant Hydrodynamic Model}

In this section we consider the non-trivial case of a locally equilibrium
system satisfying the conditions (\ref{u-n-x-conditions}). It corresponds to
the well-known and the physically important hydrodynamical solution of the
1D boost-invariant expansion \cite{Bjorken}.

Let us introduce the standard variables to analyze this hydrodynamic
solution. The space-time variables in terms of rapidity $y$ and proper time
of the expansion $\tau $ look as follows 
\begin{equation}  \label{hydr-variables}
t=\tau \cosh y,x_L=\tau \sinh y,u^0=\cosh y,u_L=\sinh y,d\sigma ^\mu =u^\mu
\tau d^2x_Tdy
\end{equation}

The system is considered on the hypersurface $\tau =const$, where the
inverse of temperature $\beta =const.$

The particle momentum can be also expressed in terms of the particle
longitudinal rapidity: 
\begin{equation}  \label{part-variables}
\begin{array}{c}
p=\left( m_p\cosh \theta _p, {\bf p}_T,m_p\sinh \theta _p\right) \quad
,\quad m_p\equiv m_T(p)=\sqrt{m^2+{\bf p}_T^2} \\ 
\\ 
d^3p=m_p\cosh \theta _pd^2p_Td\theta _p;\, \\ 
\\ 
\left[ a(p),a^{+}(p^\prime )\right] =\left[ a({\bf p}_T,\theta _p),a^{+}(%
{\bf p}_T^\prime ,\theta _{p^\prime })\right] =\delta ^2({\bf p}_T-{\bf p}%
_T^\prime )\delta (\theta _p-\theta _{p^\prime })
\end{array}
\end{equation}

To simplify the problem let us assume the transverse radius of the
hydrodynamic tube to be much larger than the hydrodynamic length $\tau $. So
one can neglect the influence of a finite transverse size of the system on
the form of spectra. The calculation of the basic functions (\ref{G}), $G$
and $\overline{G}$, that is easy to do using the variables (\ref
{hydr-variables}), (\ref{part-variables}) with an infinite transverse
region, give us the following results 
\begin{equation}
G(p,k)=-\frac{m_k\tau \beta }{2\pi \cosh \theta _k}\delta ^2({\bf p}_T-{\bf k%
}_T)\int\limits_{-\infty }^\infty dz\sqrt{z^2+1}\exp \left[ i2m_k\tau z\sinh
\left( \frac{\theta _p-\theta _k}2\right) \right]  \label{G-boost}
\end{equation}

\begin{equation}
\overline{G}(p,k)=\frac{m_k\tau \beta }{\pi \cosh \theta _k}\delta ^2({\bf p}%
_T+{\bf k}_T)\int\limits_1^\infty dz\sqrt{z^2-1}\exp \left[ -i2m_k\tau
z\cosh \left( \frac{\theta _p-\theta _k}2\right) \right]  \label{G-bar-boost}
\end{equation}

The functions (\ref{G-boost}), (\ref{G-bar-boost}) are the distributions
(the generalized functions). For example, for globally equilibrium systems $%
G(p,k)\propto \delta ^3({\bf p-k)}$. They can be considered as the Fourier
transforms of the distributions $\sqrt{1+z^2\mbox{ }}$ and $\theta (z-1)%
\sqrt{z^2-1}$ acting in the space of rapidly decreasing functions $f(\theta
) $. According to the general rules of the operations with the so-called
tempered functions \cite{Distributions}, we shall mean or directly
substitute the regular functions 
\begin{equation}
G_\epsilon =G\left( \sqrt{1+z^2}\rightarrow e^{-\epsilon \left| z\right| }%
\sqrt{1+z^2}\right) \;,\;\overline{G}_\epsilon =\overline{G}\left( \sqrt{%
z^2-1}\rightarrow e^{-\epsilon \left| z\right| }\sqrt{z^2-1}\right)
\label{G-regular}
\end{equation}

\noindent instead of (\ref{G-boost}), (\ref{G-bar-boost}) and use limit $%
\epsilon \rightarrow 0$ in the final expressions.

Let us rewrite the integral equations (\ref{integral-eqs1}) in the variables
(\ref{hydr-variables}), (\ref{part-variables}). Note that the element ${\rm K%
}_n^{ij}$ of the $n$-th iteration of the matrix kernel consists of even
number ${\rm K}^{12}$ or ${\rm K}^{21}$ in Eq. (\ref{K-n}) if $i=j$ and odd
number of them if $i\neq j$. So taking into account the structure of the
operators $G$ and $\overline{G}$, the solution of the integral equations (%
\ref{integral-eqs1}) can be presented in the form 
\begin{equation}
\left\langle a(p)a^{+}(k)\right\rangle ={\cal A}_1(\theta _p-\theta
_k)\delta ^2({\bf p}_T-{\bf k}_T),\;\left\langle a(p)a(k)\right\rangle =%
{\cal A}_2(\theta _p-\theta _k)\delta ^2({\bf p}_T+{\bf k}_T)
\label{A-theta}
\end{equation}

Then we have after the integration of Eq.(\ref{integral-eqs1}) over $d^2{\bf %
k}_T$

\begin{equation}
\sum_{n=1}^\infty \frac 1{n!}\int d\theta _k{\bf K}_n(\theta _p-\theta _k)%
{\cal A}(\theta _k-\theta _{p^{\prime }})=\left( 
\begin{array}{c}
-\delta (\theta _p-\theta _{p^{\prime }}) \\ 
0
\end{array}
\right)  \label{eqs1-theta}
\end{equation}

\noindent here the matrix ${\cal A}=({\cal A}_1,{\cal A}_2)$ is defined by
Eq.(\ref{A-theta}) and ${\bf K}_n(\theta _p-\theta _k)$ are the $n$-th
iteration of the kernel

\begin{equation}  \label{K-theta}
{\bf K}(\theta _p-\theta _k)=\left( 
\begin{array}{cc}
G(\theta _p-\theta _k) & \overline{G}(\theta _p-\theta _k) \\ 
-\overline{G}^{*}(\theta _p-\theta _k) & -G^{*}(\theta _p-\theta _k)
\end{array}
\right)
\end{equation}

\noindent The matrix elements in (\ref{K-theta}) are defined by Eqs. (\ref
{G-boost}),(\ref{G-bar-boost}):

\begin{equation}  \label{G-theta}
\begin{array}{l}
G(\theta _p-\theta _k)\delta ^2( {\bf p}_T-{\bf k}_T)=m_T\cosh \theta
_kG(p,k) \\ 
\mbox{ } \\ 
\overline{G}(\theta _p-\theta _k)\delta ^2({\bf p}_T+{\bf k}_T)=m_T\cosh
\theta _k \overline{G}(p,k)
\end{array}
\end{equation}

\noindent Fourier transform of Eq.(\ref{eqs1-theta}) gives

\begin{equation}  \label{eqs1-Fourier}
\left[ \exp {\bf K}(t)-{\bf I}\right] {\cal A}(t)=\left( 
\begin{array}{c}
-1 \\ 
0
\end{array}
\right)
\end{equation}

\noindent where {\bf I }is unit matrix, ${\bf K}(t)$ is Fourier-transformed
matrix (\ref{G-theta}).

The solution of the Eq.(\ref{eqs1-Fourier}) is easy to find after the
diagonalization of the ${\bf K(}t)$-matrix. The proper values of the
corresponding characteristic equations are

\begin{equation}
\omega _1=-\omega (t),\;\omega _2=\omega (t)\equiv \sqrt{G^2(t)-\overline{G}%
^{*}(t)\overline{G}(t)}  \label{omega}
\end{equation}

\noindent and the matrix U transforming the coordinates ${\cal A}_1{\cal \ }$%
and ${\cal A}_2$ of the vector ${\cal A}$ to the new ones ${\cal A}%
_1^{\prime {\cal \ }}$and ${\cal A}_2^\prime $ and with the diagonal matrix $%
{\bf K}$ has the form

\begin{equation}  \label{U}
U\propto \left( 
\begin{array}{cc}
(G-\omega )^2 & (G-\omega ) \overline{G} \\ 
(G-\omega )\overline{G}^{*} & (G-\omega )^2
\end{array}
\right)
\end{equation}

Now one can get the solution of the Eq.(\ref{eqs1-Fourier}). It is the
following

\begin{equation}  \label{a/a-t}
\left\langle a^{+}a\right\rangle _t={\cal A}_1-1=\frac 1{\exp \omega -1}%
\frac{(G-\omega )^2+\overline{G}^{*}\overline{G}\exp \omega }{(G-\omega )^2-%
\overline{G}^{*}\overline{G}}
\end{equation}

\begin{equation}  \label{a-a-t}
\left\langle aa\right\rangle _t={\cal A}_2=-\frac{\overline{G}^{*}(G-\omega )%
}{(G-\omega )^2-\overline{G}^{*}\overline{G}}\frac{1+\cosh \omega }{\sinh
\omega }
\end{equation}

Before we will analyze the analytical structure of the Fourier components (%
\ref{a/a-t}) and (\ref{a-a-t}) of the pair operator averages, it is
necessary to consider the principal problem. According to the construction
of the matrix density ${\bf \rho }$ (\ref{rho-alpha}) the physical vacuum
defined as $a\left| 0\right\rangle =0$ is not the proper vector of $\rho $
because of the presence of the terms proportional to $a^{+}a^{+}$ in the
energy-momentum tensor. The last term does not vanish after the integration
over $d\sigma $ except for the case $\sigma $: $t=const,$ $\beta =const$.
This means that at zero temperature, in the limit $\beta \rightarrow \infty $
($T\rightarrow 0$), the operator averages such as (\ref{a/a-t}) and (\ref
{a-a-t}) do not describe the modes of the physical vacuum $\left|
0\right\rangle $ but is associated with the modes of some ''lowest'' state $%
\left| 0^{\prime }\right\rangle $ of the operator $\int d\sigma _\nu \beta
_\mu \widehat{T}^{\mu \nu }(x)$. So all the exciting modes at finite
temperature appear over background state $\left| 0^{\prime }\right\rangle $
and we have to renormalize the averages:

\noindent 
\begin{equation}
\left\langle a^{+}a\right\rangle _t^{ren}=\left\langle a^{+}a\right\rangle
_t-\left\langle a^{+}a\right\rangle _t^{\beta \rightarrow \infty }=(\exp
\omega -1)^{-1}(1+2\left\langle a^{+}a\right\rangle _t^{\beta \rightarrow
\infty }),etc,..  \label{renorm}
\end{equation}

\noindent After this important remark we have finally

\begin{equation}
\left\langle a^{+}a\right\rangle _t^{ren}=\frac 1{\exp \omega -1}\frac{%
\left| G\right| }\omega ,\;\left\langle aa\right\rangle _t^{ren}=\frac
1{\exp \omega -1}\frac{\overline{G}^{*}}\omega ,\;\left\langle
a^{+}a^{+}\right\rangle _t^{ren}=\frac 1{\exp \omega -1}\frac{\overline{G}}%
\omega  \label{a-a-ren}
\end{equation}
where $\omega $ is defined by Eq.(\ref{omega}) and

\noindent 
\begin{equation}
\begin{array}{l}
G(t)=-\frac{m_T^2\beta \tau }\pi \int dx\sqrt{1+x^2}\int d\theta
e^{i2m_T\tau \,x\sinh \theta +\,i2t\theta } \\ 
\\ 
=-\frac{\beta \cosh (\pi t)}{\pi \tau }\int\limits_0^\infty dz\sqrt{%
(2m_T\tau )^2+z^2}K_{i2t}(z)
\end{array}
\label{G(t)}
\end{equation}
\noindent 
\begin{equation}
\begin{array}{l}
\overline{G}(t)=\mbox{ }\lim_{\epsilon \rightarrow 0}\frac{m_T^2\beta \tau }%
\pi \int d\theta e^{it\theta }\int\limits_1^\infty dz\sqrt{z^2-1}%
e^{-i2m_T\tau \,z\cosh \frac \theta 2-\epsilon \,z} \\ 
\\ 
=im_T\beta \int d\theta \frac{e^{i2t\theta }}{2\cosh \theta }%
H_1^{(2)}(2m_T\tau \cosh \theta )
\end{array}
\label{G(t)-bar}
\end{equation}

Here $K_\nu $ is the modified Bessel function and $H_1^{(2)}(z)$ is the
Hankel function of second order. The functions $G(t)$ and $\overline{G}(t)$
exhibit the following asymptotic behavior:

\begin{itemize}
\item  $m_{T}\tau \gg 1$%
\begin{equation}
\begin{array}{l}
G(t)\cong -\beta m_{T}[\sqrt{1+t^{2}/(m_{T}\tau )^{2}}+\frac{1}{24(m_{T}\tau
)^{2}}\left( 1+t^{2}/(m_{T}\tau )^{2}\right) ^{-\frac{5}{2}}\times  \\ 
\\ 
(1+\frac{3t}{m_{T}\tau }+\frac{t^{2}}{(m_{T}\tau )^{2}})+O((m_{T}\tau
)^{-4})]
\end{array}
\label{G(t)-L1}
\end{equation}
\noindent 
\begin{equation}
\left| \overline{G}(t)\right| \cong \beta m_{T}\left[ \frac{1}{2m_{T}\tau }%
\left( 1+t^{2}/(m_{T}\tau )^{2}\right) ^{-1}+O((m_{T}\tau )^{-2})\right] 
\label{G(t)bar-L1}
\end{equation}
\end{itemize}

\noindent 
\begin{equation}  \label{omega(t)-L1}
\omega (t)\cong \beta m_T\left( 1+t^2/(m_T\tau )^2-\frac{1-3t/2m_T\tau
-t^2/2m_T\tau }{6(m_T\tau )^2\left( 1+t^2/(m_T\tau )^2\right) ^2}\right)
^{\frac 12}
\end{equation}

\begin{itemize}
\item  $m_{T}\tau \ll 1,\;t\leq m_{T}\tau $

\noindent 
\begin{equation}
G(t)\cong -\frac{t\beta \coth \pi t}{\tau }\left\{ 1+\frac{(m_{T}\tau )^{2}}{%
2}\left[ \ln {}^{2}\frac{m_{T}\tau }{2}+\ln \frac{m_{T}\tau }{2}+1\right]
\right\} +O\left( (m_{T}\tau )^{4}\right)   \label{G(t)-S1}
\end{equation}

\noindent 
\begin{equation}
\begin{array}{l}
\overline{G}(t)\cong -\frac{t\beta }{\tau \sinh \pi t}\left\{ 1+\frac{%
(m_{T}\tau )^{2}}{2}\left[ \ln {}^{2}\frac{m_{T}\tau }{2}+\ln \frac{%
m_{T}\tau }{2}+1-\pi ^{2}\right] \right\} + \\ 
\\ 
+i\frac{\beta (m_{T}\tau )^{2}}{2\tau }\ln \frac{m_{T}\tau }{2}+O\left(
(m_{T}\tau )^{4}\right) 
\end{array}
\label{G(t)bar-S1}
\end{equation}

\noindent 
\begin{equation}
\omega (t)\cong \beta m_{T}\sqrt{1+t^{2}/(m_{T}\tau )^{2}}
\label{omega(t)-S1}
\end{equation}
\end{itemize}

The asymptotics is found by the saddle-point method for $m_T\tau \gg 1$ and
using the representation of Eqs.(\ref{G(t)}), (\ref{G(t)-bar}) as the sum of
the hypergeometrical functions (see \cite{Prudnikov}) for $m_T\tau \ll 1$.

It is important to note that function $\omega $ is described by the
expression (\ref{omega(t)-S1}) for all values $m_T\tau $ with the accuracy
more than, at least, one percent as the numerical calculations have shown.
It means, that the main factor $\left( \exp \omega -1\right) ^{-1}$ in Eqs.(%
\ref{renorm}) is actually the Bose-Einstein function (\ref{BE-distr}). We
believe that there is deep analytical foundation of this fact but we cannot
explain this mathematical gift now.

It is worth to mention that the introduction of a non-zero chemical
potential associated with violation of the chemical equilibrium in boson gas
(overpopulation) leads to changing of all results by the substitution: $%
G(t)\rightarrow G(t)-\mu .$

Let us calculate the single particle spectrum for the case $\mu =0$. The
pair operator average is described by the inverse Fourier transform of Eq.(%
\ref{a-a-ren}). With Eq.(\ref{A-theta}) taken into account we have\noindent 
\begin{equation}
\left\langle a^{+}(p)a(p^{\prime })\right\rangle =\delta ^2({\bf p}_T-{\bf p}%
_T^{\prime })\frac 1{2\pi }\int dte^{-i(\theta _p-\theta _{p^{\prime
}})t}\;\left\langle a^{+}a\right\rangle _t  \label{a/a-p}
\end{equation}

\noindent 
\begin{equation}  \label{a-a-p}
\left\langle a(p)a(p^\prime )\right\rangle =\left\langle
a^{+}(p)a^{+}(p^\prime )\right\rangle ^{*}=\delta ^2({\bf p}_T+{\bf p}%
_T^\prime )\frac 1{2\pi }\int dte^{-i(\theta _p-\theta _{p^{\prime
}})t}\;\left\langle aa\right\rangle _t
\end{equation}

One can calculate the formal distribution function defined by (\ref
{wigner-def}) using the variable $\zeta =$Ar$\sin $h$\frac{v_L}{2m_T\cosh
\theta }$ and the representation (\ref{part-variables}) for momenta.

\begin{equation}  \label{wigner-hydr}
\begin{array}{l}
f(x,p)_\sigma =\frac 2{(2\pi )^4}\int d\zeta e^{i2m_T\tau \sinh (\theta
-y)\sinh \zeta }\int dt\,e^{-i2t\zeta }\left\langle a^{+}a\right\rangle _t
\\ 
\\ 
\approx \frac{\left\langle a^{+}a\right\rangle (t=m_T\tau \sinh (\theta -y)) 
}{(2\pi )^3}
\end{array}
\end{equation}

The hydrodynamics rapidity $y$ defines the position of point $x$ according
to Eq.(\ref{hydr-variables}). The last approximation in Eq.(\ref{wigner-hydr}%
) is valid when the Fourier transform $F\left[ \left\langle
a^{+}a\right\rangle _t\right] (\zeta )$ is fast decreasing in $\zeta $, that
is satisfied. So in the local rest system of the fluid element, $y=0$, the
result has the form

\begin{itemize}
\item  $m_T\tau \gg 1$\noindent 
\begin{equation}
\frac{d^6N}{d^3xd^3p}_{|{\bf u=0}}=f(x,p)_{\sigma |{\bf u=0}}\approx \frac{%
(2\pi )^{-3}}{\exp (\beta p_0)-1}\left( 1+\frac{m_T^2}{24\tau ^2p_0^4}\right)
\label{wigner-hydr-L1}
\end{equation}

\item  $m_T\tau \ll 1$\noindent 
\begin{equation}
\frac{d^6N}{d^3xd^3p}_{|{\bf u=0}}=f(x,p)_{\sigma |{\bf u=0}}\approx \frac{%
(2\pi )^{-3}}{\exp (\beta p_0)-1}\frac 1{\pi \tau p_0}=f_{BE}(p)\left(
2f_M(p)\right)  \label{wigner-hydr-S1}
\end{equation}
\end{itemize}

We introduce here the designation

\begin{equation}  \label{Milne-distr}
f_M(p)=\left\langle a^{+}a\right\rangle _{t=m_T\tau \sinh \theta }^{\beta
\rightarrow \infty }+\frac 12=\left( e^{2\pi \tau p_0}-1\right) _{|\tau
p_0\ll 1}^{-1}
\end{equation}

The function $f_M(p)$ describes the ''heat'' spectra with the
''temperature'' $T_{eff}=1/2\pi \tau $ and the real temperature $T=0$ for
the so-called Milne's particles \cite{Birrel}, which appear when the
2-dimensional field system is quantized in the hyperbolic space-time that is
known as Milne's Universe \cite{Milne}\footnote{%
The same spectrum with $\tau =1/a$ is well known also for the Rindler's
particles that appear in uniformly accelerating reference system with
acceleration $a$ due to the quantization on a time-like hyperboloid formed
by the world lines of accelerating observers.}. It happens due to the mixing
of the positive and negative frequency components of a field in the
hyperbolic world in compare with the Minkovski one. In the boost-invariant
hydrodynamic model this hyperbolic space-time is formed by the isotherms.
The state of the ''lowest'' energy for the operator $\int d\sigma _\nu u_\mu 
\widehat{T}^{\mu \nu }$ containing in the statistical operator ${\bf \rho }$
was found in the Ref.\cite{Sommerfield} in the limit corresponding to $\tau
\rightarrow 0$ in our case. So the unrenormalized spectrum $\left\langle
a^{+}a\right\rangle _t^{\beta \rightarrow \infty }$ gives the standard
result (\ref{Milne-distr}) for the spectrum of the Milne's particles in this
limit. After the background subtraction procedure, which is necessary since
we study the particles against the background of the Minkovski vacuum but
not the Milne's one, the trace of this phenomenon can be observed at the
finite temperature only as it is demonstrated by the Eqs.(\ref{renorm}), (%
\ref{wigner-hydr-S1}). The physical reason for this lies in the creation of
additional quanta of a bosonic field due to the interference of the positive
and negative frequency components of a field when the latter begins to
embrace the ''hyperbolically'' expanding medium. Naturally this effect may
be noticeable only if the wave-length of the quanta is much large than the
length of homogeneity , in other words, when non-localized quanta ''feels''
a space-time inhomogeneity of a medium.

The numerical demonstration of this effect for the distribution function is
represented in Fig.1.

\begin{figure}[tbh]
\vspace*{-1.0cm} \insertplot{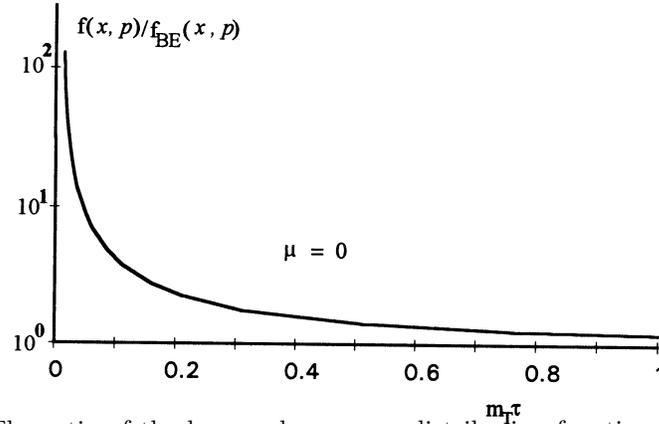} \vspace*{-1.9cm}
\caption{ The ratio of the boson phase-space distribution functions in
expanding matter to the Bose-Einstein thermal distribution. }
\label{fig1}
\end{figure}

Note, that at $\mu \neq 0$ there is noticeable deviation of the distribution
from the Bose-Einstein result even at $m_T\tau \gg 1$ that began to be very
large when $\mu \rightarrow \mu _{cr}=m(1-1/2m\tau ):$

\begin{equation}
\frac{f(p,x)}{f_{BE}(p,x)}\rightarrow \left( 1-\frac{m_T^2}{(2\tau
p_0(p_0-\mu ))^2}\right) ^{-1}  \label{chemical}
\end{equation}
The single particle spectra according to (\ref{spectra-def}),(\ref{a/a-p})
for large enough radius $R$ of the hydrodynamic tube has the form

\begin{equation}  \label{Mt-spectra-hydr}
\frac{d^2N}{m_Tdm_Td\theta }=\frac{\pi R^2}{(2\pi )^3}\int dt\left\langle
a^{+}a\right\rangle _t
\end{equation}

The effects of inhomogeneity will be important at the early stage, $\tau $$%
\approx 1$ fm for low-mass gluons or quasi-bosons if such objects there are
at this stage. It could lead to the enhancement of photons and dileptons
with low invariant mass. At the final freeze-out stage, even if $m_T\tau \gg
1$, the distortion of spectra due to the inhomoheneity will take place when
the violation of the chemical equilibrium in boson gas is strong: the
chemical potential $\mu $ is closed to $\mu _{cr}$.

Note, however, that we did not take into account a longitudional geometrical
size of the system, $R_L.$ It means that our consideration is limited by the
condition $p^0\geq 1/R_L.$ The condition $p^0\gg 1/R_T$ have been supposed
before. Note, that as it was found in \cite{Matrahaza}, a smallness of
transverse geometrical size of the system, $R_T\leq 1/p^0,$ leads to an
reduction of number of soft quanta in locally equilibrium system in
comparision with the Bose-Einstein spectra at the same values of
thermodynamic parameters.

\section{Two-Particle Correlations}

The generalized Wick's theorem (\ref{Wick-theorem}) allows one to consider
many-particle inclusive spectra and correlations for concrete fields. Let us
begin from pion interferometry. The isotopic invariant Lagrangian of free
pion field in the representation of real 3-component pion field operator $%
\pi $ and nonzero commutators have the form ($\;i,j=1,2,3.$):\noindent 
\begin{equation}
{\cal L}_{{\bf \pi }}=\frac 12\left( {\bf \pi }_{;\mu }{\bf \pi }^{;\mu
}\right) -\frac{m^2}2({\bf \pi \cdot \pi ),\;}\left[ \dot{\pi}_j(x^0,{\bf x}%
),\pi _j(x^0,{\bf x^{/}})\right] =\frac 1i\delta _j^i\delta ({\bf x-x^{/}})
\label{iso-lagrangian}
\end{equation}
\ So it is easy to see that dynamic equations split into three independent
ones for each real component $\pi _j$; the statistical operator ${\bf \rho }$
is the product of three commuting exponents and all the previous results are
preserved for each Hermitian field $\pi _j\Longleftrightarrow \varphi $
independently.

As it well known, the creation and annihilation operators of pions $\pi
_{+},\pi _{-},\pi _0$ are described by the pion field in the complex
representation and are connected with corresponding operators of the
Hermitian ($\pi _1,\pi _2,\pi _{3\mbox{ }}$)-field in the following manner 
\cite{Bjorken-Drell}

\begin{equation}  \label{a-a-pi}
\begin{array}{ccc}
a_{+}=\frac 1{\sqrt{2}}\left( a_1+ia_2\right) & a_{+}^{+}=\frac 1{\sqrt{2}%
}\left( a_1^{+}-\;ia_2^{+}\right) & a_0^{+}=a_3^{+} \\ 
a_-=\frac 1{\sqrt{2}}\left( a_1-\;ia_2\right) & a_-^{+}=\frac 1{\sqrt{2}%
}\left( a_1^{+}+\;ia_2^{+}\right) & a_0=a_3
\end{array}
\end{equation}

Using the Wick's theorem (\ref{Wick-theorem}) and Eq.(\ref{iso-lagrangian})
we express the inclusive spectra (\ref{spectra-def}) through the results (%
\ref{a/a-p}),(\ref{a-a-p}),(\ref{a-a-ren}) for real scalar field taking into
account that 
\begin{equation}  \label{a-a-ij}
\left\langle a_i^{+}a_j\right\rangle =\delta _i^j\left\langle
a^{+}a\right\rangle ,\,\left\langle a_i^{+}a_j^{+}\right\rangle =\delta
_i^j\left\langle a^{+}a^{+}\right\rangle ,\;\left\langle
a_i\,a_j\right\rangle =\delta _i^j\left\langle a\,a\right\rangle
\end{equation}
The results are

\begin{itemize}
\item  for $\pi ^{-}\pi ^{-}$ (and similarly $\pi ^{+}\pi ^{+}$ ) pion pairs 
\begin{equation}
\begin{array}{r}
\left\langle a_{+}^{+}(p_1)a_{+}^{+}(p_2)a_{+}(p_1)a_{+}(p_2)\right\rangle
=\left\langle a^{+}(p_1)a(p_1)\right\rangle \left\langle
a^{+}(p_2)a(p_2)\right\rangle \\ 
\\ 
+\left\langle a^{+}(p_1)a(p_2)\right\rangle \left\langle
a^{+}(p_2)a(p_1)\right\rangle
\end{array}
\label{pi/pi/}
\end{equation}

\item  for $\pi ^{+}\pi ^{-}$ pairs 
\begin{equation}
\begin{array}{r}
\left\langle a_{+}^{+}(p_1)a_{-}^{+}(p_2)a_{+}(p_1)a_{-}(p_2)\right\rangle
=\left\langle a^{+}(p_1)a(p_1)\right\rangle \left\langle
a^{+}(p_2)a(p_2)\right\rangle \\ 
\\ 
+\left\langle a^{+}(p_1)a^{+}(p_2)\right\rangle \left\langle
a(p_2)a(p_1)\right\rangle
\end{array}
\label{pi/pi}
\end{equation}

\item  for $\pi ^0\pi ^0$-pairs 
\begin{equation}
\begin{array}{c}
\left\langle a_0^{+}(p_1)a_0^{+}(p_2)a_0(p_1)a_0(p_2)\right\rangle
=\left\langle a^{+}(p_1)a(p_1)\right\rangle \left\langle
a^{+}(p_2)a(p_2)\right\rangle \\ 
\\ 
+\left\langle a^{+}(p_1)a(p_2)\right\rangle \left\langle
a^{+}(p_2)a(p_1)\right\rangle +\left\langle
a^{+}(p_1)a^{+}(p_2)\right\rangle \left\langle a(p_2)a(p_1)\right\rangle
\end{array}
\label{pi0-pi0}
\end{equation}
\end{itemize}

Such a structure of the two pion spectra has been gotten in the Ref.\cite
{Weiner} in the quantum optics model (Gaussian random sources) and in the
Ref.\cite{Sin-Tolst} in two particle quantum mechanical approach. The
significance of our result is that we calculate the averages $\left\langle
a^{+}a\right\rangle ,\,\left\langle a^{+}a^{+}\right\rangle $ and$%
\;\left\langle a\,a\right\rangle $ in the concrete realistic model of
locally equilibrium hadron medium. The correlation function $C(p_1,p_2)$ for
different pion pairs can be obtained by the division of the two particle
spectra (\ref{pi/pi/})-(\ref{pi0-pi0}) by the single particle momentum
distributions (\ref{a-a-ij}) of the corresponding pions. The theoretical
maximum of the correlation functions $C(p_1,p_2)$ is achieved at ${\bf p}_1=%
{\bf p}_2=0$ and can be easily estimated using the saddle point method: 
\begin{equation}
\begin{array}{c}
\max C(\pi ^{\pm }\pi ^{\pm })=2,\quad \max C(\pi ^{\pm }\pi ^{\mp })\approx
1+\left| \frac{\overline{G}(t=0)}{G(t=0)}\right| ^2, \\ 
\\ 
\;\max C(\pi ^0\pi ^0)\approx 2+\left| \frac{\overline{G}(t=0)}{G(t=0)}%
\right| ^2
\end{array}
\label{corr-max}
\end{equation}
According to the asymptotic expansion (\ref{G(t)-L1})-(\ref{G(t)bar-S1}) 
\begin{eqnarray}
\left| \frac{\overline{G}(t=0)}{G(t=0)}\right| ^2 &\simeq &\frac{m\beta }{%
(2m\tau )^3}\mbox{\ \ for\ \ }m\tau \gg 1  \nonumber  \label{Gbar-G-ratio} \\
&&\mbox{and}  \label{Gbar-G-ratio} \\
\mbox{\ \ \ }\left| \frac{\overline{G}(t=0)}{G(t=0)}\right| ^2 &\simeq
&1-(\pi m\tau )^2\mbox{\ \ for\ \ }m\tau \ll 1  \nonumber
\end{eqnarray}

The corresponding plot of $\max C(m\tau )$ is demonstrated in Fig. 2.

\begin{figure}[tbh]
%\insertplot{fig2sin.ps} 
\epsfxsize=10cm\epsffile{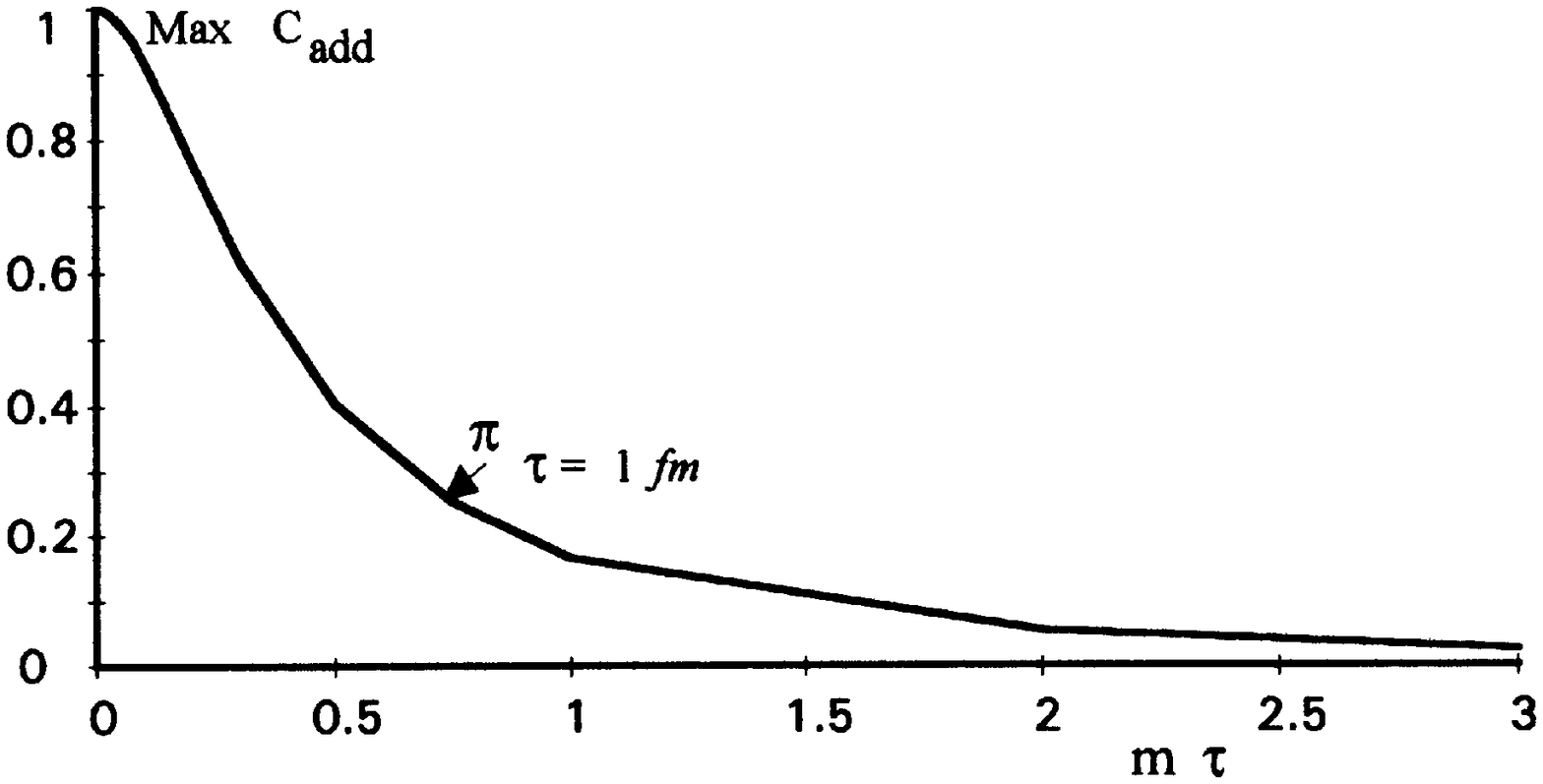}
\caption{ The aditional contribution to the correlation peak value for
neutral and oppositely charged bosons at $\mu =0$. }
\label{fig2}
\end{figure}

For pions this value is noticeable only at $\tau \approx $1 fm and is 0.25
approximately, so $\max C_{\pi ^0\pi ^0}(\tau =1$fm$)\approx 2.25,$ $\max
C_{\pi ^{-}\pi ^{+}}(\tau =1$fm$)\approx 1.25$. These values differ from the
quantum-mechanical results \cite{Sin-Tolst} where $\max C_{\pi ^0\pi ^0}=2$,
and $C_{\pi ^{-}\pi ^{+}}\equiv 1.$ This means that the effect has
relativistic nature (decomposition of a field into the positive and negative
frequency components and their interference in finite regions of
homogeneity).

Note that if the chemical potential $\mu $ tends to critical value $\mu _c$
then even for large $m\tau \gg 1$ we have for maximal intercept (when the
both quanta are very soft, ${\bf p}_1\simeq {\bf p}_2\simeq 0$) unusually
large values: $\simeq 2$ for $\pi ^{+}\pi ^{-}$ and $\simeq 3$ for $\pi
^0\pi ^0$ . For equally charged pions $\pi ^{+}\pi ^{+}$ and $\pi ^{-}\pi
^{-}$ the intercept has the standard value =2 for any $m\tau $ and $\mu $.

In the typical experimental situation when $m\tau \gg 1$ and $\mu =0$ the
Bose-Einstein correlation functions have the standard structure (\ref
{double1}) for all sorts of identical pions and longitudinal projection of
the correlation function can be approximated by the expression

\begin{equation}
C(p,q_L;{\bf q}_T={\bf 0})\approx 1+\frac{\exp \left[ \frac 2{\lambda
_L^2}\left( 1-\sqrt{1+\tau ^2\lambda _L^4q_L^2}\right) \right] }{\left(
1+\tau ^2\lambda _L^4q_L^2\right) ^{3/2}}\stackrel{p_T\rightarrow \infty }{%
\longrightarrow }1+\exp \left[ -\tau ^2\lambda _L^2q_L^2\right]
\label{corr-hydr}
\end{equation}
where $\lambda _L\approx \sqrt{\frac T{m_T}}.$ Here $T$ is the ''freeze
out'' temperature corresponding to proper time $\tau $ when the particles
leave the expanding matter. This result is obtained by the saddle-point
method from (\ref{pi/pi/}) and has the asymptotic form for longitudinal
interferometry radius firstly obtained in Refs. \cite{Sin-qm88}, \cite
{Makhl-Sin} for $\beta m_T\gg 1.$

The two- particle spectra for charge $K^{+},K^{-}$-kaons are described by
Eqs. (\ref{pi/pi/}), (\ref{pi/pi}) with substitution $\pi ^{\pm }\rightarrow
K^{\pm }$ because the complex representation for these fields can be
replaced by the real one in the same manner as for charge pions. The same
concerns of $K^0,\overline{K}^0$-pairs. Because of relatively large kaon
mass the role of the addition term in the correlation function $%
C(K^{+},K^{-})$ is negligible. For the correlation functions of identical
kaons asymptotic form in Eq.(\ref{corr-hydr}) can be used in all momentum
region.

The correlations in expanding photon gas are described by the formula (\ref
{pi0-pi0}) with multiplier $1/2$ at the second and third terms arising due
to random polarization of photons. In this case the additional third term
gives a good contribution for very soft photons producing $\max C=2$ instead
of =1.5 without the third term.

\section{Conclusions}

In this paper we give the theoretical analyses of spectra and correlations
in inhomogeneous weekly interacting boson gas. For the purpose the method of
locally equilibrium statistical operator used to calculate the averages such
as $\left\langle a^{+}(p_1)...a(p_j)...\right\rangle $ has been developed.
The problem was reduced to the system of the integro-differential and
integral equations solved analytically for physically significant model of
hydrodynamic boost-invariant expansion. The main results are:

\begin{itemize}
\item  the deviation of the particle phase-space density distribution from
the Bose-Einstein one even in main approximation neglecting dissipative
phenomena;

\item  the appearance of the additional terms in the correlation functions
of like and unlike (oppositely charged) particles as compared with the
results of the nonrelativistic quantum mechanical approach.
\end{itemize}

These effects are essential at small values $\tau p_0\leq 1$ or/and at large
enough chemical potentials $\mu \rightarrow \mu _c$. Under this condition
the ''effective'' wave-length of the quanta, ($p_0-\mu $)$^{-1}$ is larger
than the length of homogeneity $\tau $ in thermalized medium and quanta
begin to ''feel'' the all expanding matter. If $\mu =0,$ it results in the
additional number of soft quanta due to interference of positive and
negative frequency components of the relativistic quantum field in finite
regions of homogeneity. In special case of Bjorken boost-invariant picture
this effect is described by the spectrum $f_M(p_0)$ of the so-called Milne's
particles that appear at zero temperature in the hyperbolic space-time due
to a mixing of the positive and negative frequency field components
relatively to the Minkovski space. In the hydrodynamic picture the role of
the Milne's Universe is played by the expanding thermalized matter
''forming'' this hyperbolic world by the isoterms. Here there are no effects
at zero temperature: the spectrum is $f(p_0)=f_{BE}(p_0)(2f_M(p_0))$. Note
that Milne's spectrum $f_M\propto \left| \left\langle aa\right\rangle
\right| ^2$. The last value is responsible for the additional terms in the
correlation functions. Therefore, the both effects have the common nature.
They are connected with the space-time inhomogeneity of systems.

The goal of the interferometry analysis in $A+A$ collisions is to study the
space-time evolution of the matter or, roughly speaking, to find proper time 
$\tau $ of expansion. This is, actually, the average longitudinal length of
homogeneity. Generally speaking, {\it the} {\it ''interferometry
microscope'' measures the size and shape of homogeneity regions in radiating
sources}. As known, the relative smallness of the effective emitting region
is the basic condition for interferometry method to be applied
experimentally. Under this circumstance the taking into account of the new
effects for spectra and correlations of effectively soft bosons, ($p_0-\mu $)%
$^{-1}\gg \tau $ become to be important.

The experimental consequences of the effect in the ultra-relativistic
nucleus-nucleus collisions could also concern of the particles with small
mass such as chiral quarks, gluons, photons, etc. The theory predicts the
essential enhancement for a number of particles with small effective energy
at an early stage of the matter expansion. This can lead to the increase in
the number of photons and dileptons with small transverse momenta or small
invariant masses if they are produced in the collisions of particles with
small effective mass.

{\bf Comments}

The paper is minor modified version of the unpublished preprint
Yu.M.Sinyukov, ITP-93-8E, Kiev, 1993. Here was done firstly the
interpretation of the HBT radii as the lengths of homogeneity in radiating
systems. The modification takes into account some of the later results
published in proceedings of the conferences:

Yu.M.Sinyukov, Nucl. Phys. A566 (1994) 589c (QM 93);

Yu.M.Sinyukov. Spectra and correlations in small inhomogenious systems. In:
Hot Hadronic Matter. Theory and Experiment, (J.Letessier, H.H.Gutbrod,
J.Rafelski, eds.) p. 309, Plenum Publ., 1995. (NATO Workshop, Divonne-94);

Yu.M.Sinyukov, S.V.Akkelin, A.Yu.Tolstykh, Nucl.Phys. A610 (1996) 278c (QM
96);

Yu.M.Sinyukov, S.V.Akkelin, R.Lednicky, In Proc. of the 8th International
Workshop on Maltiparticle Production in Matrahaza (T.Csorgo et al, eds),
p.66, World Scientific, 1998.

and in the paper Yu.M.Sinyukov, B.Lorstad, Z.Phys.C61 (1994) 587. \vskip %
10pt \noindent {\large {\bf Acknowledgement}}{\normalsize \vskip 10pt
\noindent }I express my sincere thanks to V.A.Averchenkov for his assistance
in the computer calculations and S.V.Akkelin for fruitful disscussions.

{\normalsize \itemsep -1mm %\begin{thebibliography}{99} 
\parindent=8truemm \itemsep -1mm }

{\normalsize \vfill\eject
}

\end{document}